\documentclass{emulateapj} 
\usepackage{apjfonts}
\usepackage{mathrsfs}
\begin{document}
\slugcomment{To appear in ApJS}
\newcommand{\LeeZinn}{\mathcal{L}}









\shortauthors{C. Cort\'es, M. Catelan} 
\shorttitle{The RRL PLC and PCC Relations in the Str\"omgren System}

\title{The RR Lyrae Period-Luminosity-(Pseudo-)Color and Period-Color-(Pseudo-)Color 
       Relations in the Str\"omgren Photometric System: Theoretical Calibration}

\author{C. Cort\'es\altaffilmark{1,}\altaffilmark{2}}

\author{M. Catelan\altaffilmark{2}}

\altaffiltext{1}{Departamento de F\'{i}sica Te\'orica e Experimental, Universidade Federal 
do Rio Grande do Norte, Campus Universit\'ario,  59072-970 Natal, RN, Brasil; 
e-mail: cristian@dfte.ufrn.br}

\altaffiltext{2}{Departamento de Astronom\'{i}a y Astrof\'{i}sica, Pontificia Universidad
Cat\'olica de Chile, Avenida Vicu\~na Mackena 4860, 782-0436 Macul, Santiago, Chile; 
e-mail: mcatelan@astro.puc.cl}

\begin{abstract}
We present a theoretical calibration of the RR Lyrae period-luminosity-color  
and period-color-color 
relations in the multiband \emph{uvby} Str\"omgren photometric system. Our theoretical 
work is based on calculations of synthetic horizontal branches (HBs) for 
four different metallicities, fully taking into account evolutionary 
effects for a wide range in metallicities and HB morphologies. While 
our results show that ``pure'' period-luminosity and period-color 
relations do not exist in the Str\"omgren system, which is due to the large 
scatter that is brought about by evolutionary effects when the $uvby$ bandpasses 
are used, they also reveal that such scatter can be almost completely 
taken into account by incorporating Str\"omgren pseudo-color 
[$C_0 \equiv (u\!-\!v)_0 - (v\!-\!b)_0$] terms into those equations, thus leading 
to tight period-luminosity-{\em pseudo}-color (PLpsC) and period-color-{\em pseudo}-color 
(PCpsC) relations. We provide the latter in the form of analytical fits, so that they 
can be applied with high precision even in the case of field stars. In view of the 
very small sensitivity of $C_0$ to interstellar reddening, our PLpsC and PCpsC 
relations should be especially useful for the derivation of high-precision distance 
and reddening values. In this sense, we carry out a first application of our 
relations to field RR Lyrae stars, finding evidence that the stars RR Lyr, 
SU Dra, and SS Leo~-- but not SV Hya~-- are somewhat overluminous (by amounts 
ranging from $\simeq 0.05$ to 0.20~mag in $y$, and thus $V$) with respect to 
the average for other RR Lyrae stars of similar metallicity.   
\end{abstract}


\section{Introduction}\label{sec:INTRO}
Due to the special characteritics of the \citet{bs63} $uvby$ passband system,  
it represents an invaluable tool in the study of the physical parameters of 
stars, such as effective temperature, surface gravity, metallicity, and 
even age \citep[e.g.,][]{sn89,fgea00}. More recently, 
it has also been shown that this system~-- and its \emph{u} passband in 
particular~-- provides us with a sensitive diagnostic of radiative levitation and 
gravitational settling phenomena taking place in hot horizontal branch (HB) 
stars \citep{fgea99}. While observations in this system have 
traditionally been limited to bright and nearby stellar systems, 
over the past decade and a half, with the advent 
of modern CCD detectors and increasingly large collector areas, the range of 
systems within reach of observations in the \emph{uvby} passband has been 
increasing dramatically, thus giving a renewed impetus for astrophysical 
applications of $uvby$ observations. Accordingly, the main purpose of the 
present paper is to present the first extensive theoretical 
calibration of the RR Lyrae (RRL)
period-luminosity-color and period-color-color relationships in the 
Str\"omgren system, thus enabling the latter's use for distance 
and reddening determinations throughout the old components of the Local Group.


\begin{figure*}[t]
\centering
  \includegraphics*[width=16cm]{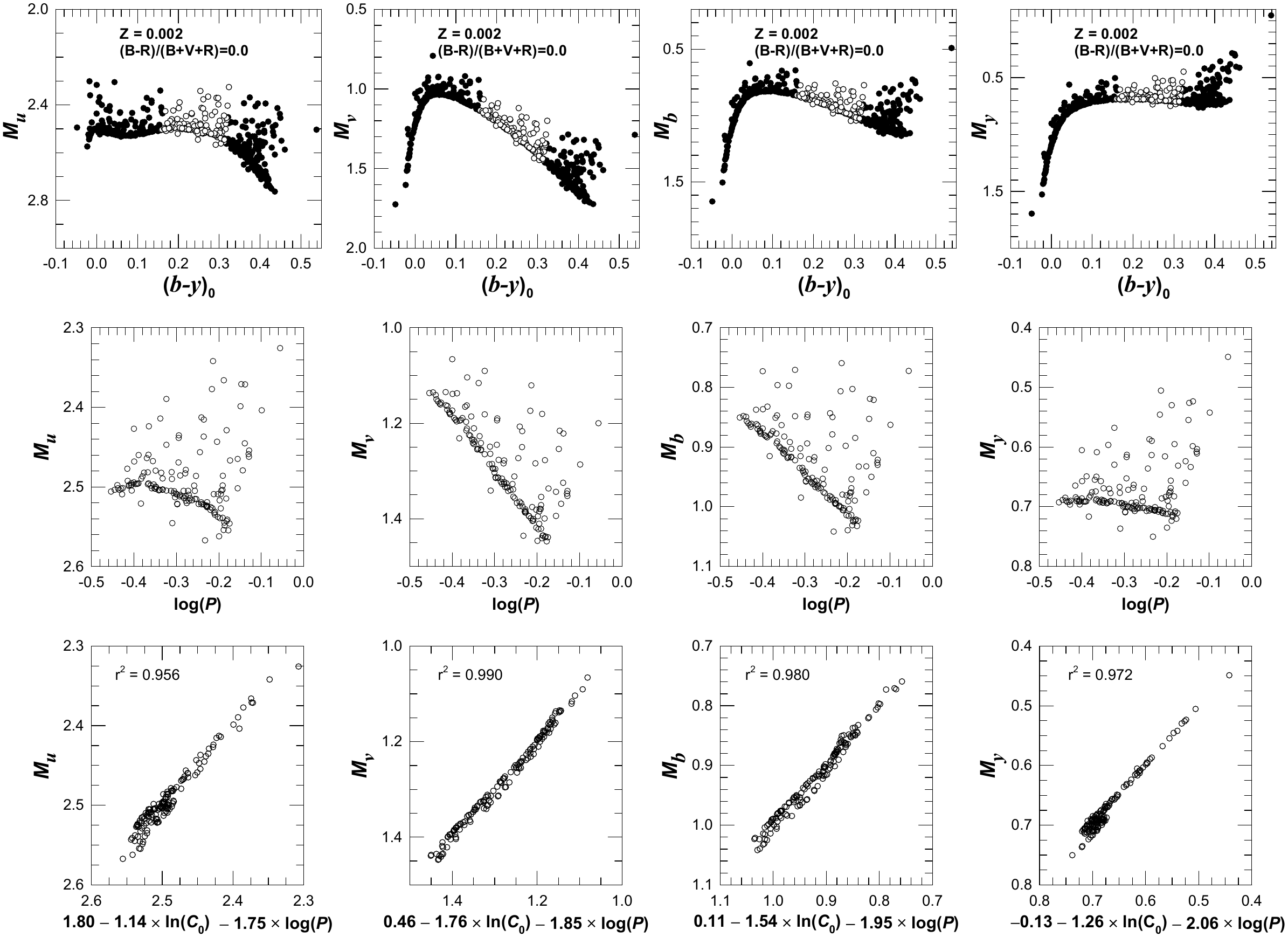}
  \caption{{\em Upper panels}: Morphology of the HB in the CMD when different 
     Str\"omgren bandpasses (from left to right, $u$, $v$, $b$, $y$) are used,   
     based on synthetic calculations. RRL variables are shown as open circles, 
     and non-variable stars as black circles. {\em Middle panels}: 
     Corresponding distributions in the absolute magnitude--log-period 
     (or PL) plane. {\em Lower panels}: Corresponding RRL
     distributions in the absolute magnitude--log-period--(pseudo-)color 
     plane. Note the dramatic reduction in scatter that is brought about with the 
	 inclusion of a $C_0$-dependent term (the correlation coefficient $r$ is shown 
	 in the lower panels). All plots 
     refer to an HB simulation with $Z = 0.002$ and an intermediate HB type, 
     as indicated in the upper panels.
   }
      \label{fig:GENMAG}
\end{figure*}


\begin{figure*}[t]
\centering
  \includegraphics*[width=13.5cm]{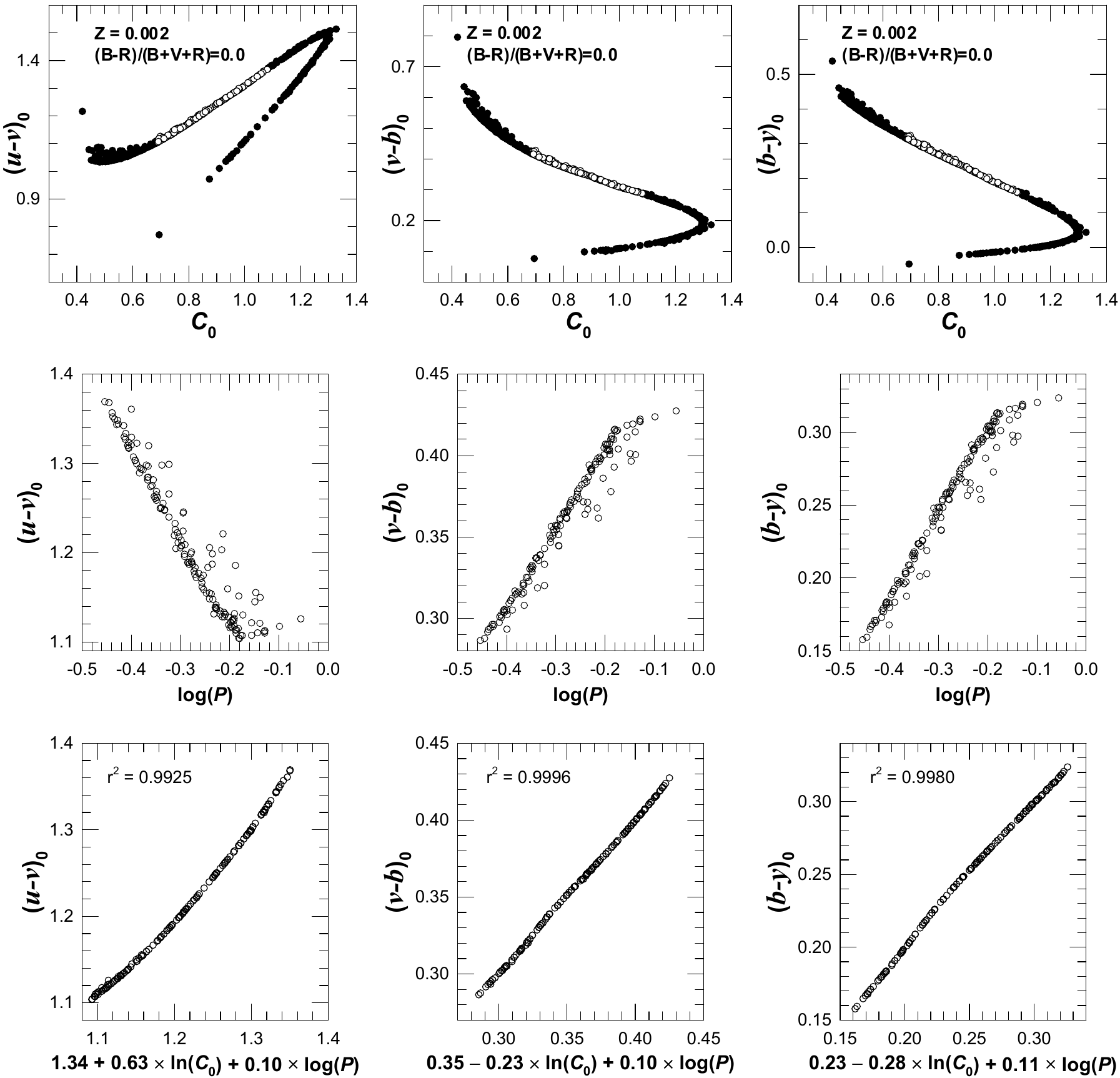}

  \caption{{\em Upper panels}: Morphology of the HB in the color-pseudo color
     plane when different Str\"omgren-based colors are used 
     [{\em left}: $(u\!-\!v)_{0}$; {\em middle}: $(v\!-\!b)_{0}$; 
     {\em right}: $(b\!-\!y)_{0}$], based on synthetic calculations. 
     RRL variables are shown as open circles, 
     and non-variable stars as black circles. {\em Middle panels}: 
     Corresponding distributions 
     in the color--log-period (or PC) plane.
     {\em Lower panels}: Corresponding RRL
     distributions in the color--log-period--(pseudo-)color plane. Note the 
	 dramatic reduction in scatter that is brought about with the 
	 inclusion of a $C_0$-dependent term (the correlation coefficient $r$ is shown 
	 in the lower panels).  All plots 
     refer to an HB simulation with $Z = 0.002$ and an intermediate HB type, 
     as indicated in the upper panels.  
      }
      \label{fig:GENCOL}
\end{figure*}

HB stars form a prop to estimate parameters in globular clusters (GC) and 
nearby galaxies, including their distance, age, and chemical composition. 
RRL variables, in particular, as the cornerstone 
of the Population~II distance scale, can help us determine the distances 
to old and sufficiently metal-poor systems, in which this type of variable 
star is commonly found in large numbers. RRL stars are radially pulsating 
variable stars with periods in the range between about 0.2~d and 1.0~d, 
and they are abundantly present in GCs and the dwarf galaxies in the 
neighborhood of the Milky Way \citep[e.g.,][ and references therein]{mc04,mc05}. 
RRL stars have also been positively identified in the M31 field 
\citep[e.g.,][]{tbea04,adea04} and in some of Andromeda's companions 
\citep*[e.g.,][]{bpea05a}, and in at least four M31 globular clusters 
\citep{gcea01}.  

While period-luminosity (PL) relations in the near-infrared bandpasses have 
been known and studied for a long time now 
\citep*[e.g.,][]{alea86,gbea01,mcea04,dpea06},  
to the best of our knowledge no such studies have ever been carried out in 
the \citet{bs63} passband system, perhaps in view of the fact that the latter 
does not contain any bandpasses in the near infrared, and indeed in the 
Johnson-Cousins system there are no good PL relations in the visual, except  
perhaps in $I$ \citep{mcea04}. However, given that the \citet{bs63} has already 
proved superior to the wide-band Johnson-Cousins system in deriving stellar 
physical parameters, it seemed to us well worth the while to 
carry out a theoretical investigation of the PL (and period-color, or PC) 
relation in this system. Accordingly, we have carried out, 
on the basis of evolutionary models and HB simulations, the first analysis 
of the subject, discovering the presence of tight PL-pseudo-color (PLpsC) 
and PC-pseudo-color (PCpsC) relations for RRL stars~-- where the pseudo-color 
is defined as $C_0 \equiv (u\!-\!v)_0 - (v\!-\!b)_0$, and is well known to 
be quite insensitive to reddening \citep[e.g.,][]{cm76}. 

We begin by presenting, in \S\ref{sec:MODELS}, the theoretical framework upon which our 
study is based. In \S\ref{sec:GENESIS}, we explain the origin of the derived PLpsC and PCpsC 
relations in the Str\"omgren system. In \S\ref{sec:CALIBRE} we provide the first 
calibration of the RRL PLpsC and PCpsC relations. \S\ref{sec:CAVEATS} 
presents some comments and warnings regarding the application of these relations 
to the analysis of empirical data. A first comparison with the 
observations is presented in \S\ref{sec:DATA}. Finally, some concluding remarks 
are provided in \S\ref{sec:CONCL}.

\section{Models}\label{sec:MODELS}
The HB simulations computed in the present paper are similar to those described 
in \citet{mc04} and \citet{mcea04}, to which the reader is referred for 
further details and references about the HB synthesis method. \citeauthor{mcea04}
argue that these models are consistent with a distance modulus to the Large 
Magellanic Cloud of $(m\!-\!M)_0 = 18.47$~mag. Note, however, that this value 
is based on the empirical prescriptions for the LMC by \citet{rgea04}; using the 
independent measurements by \citet{jbea04}, a distance modulus of 
$(m\!-\!M)_0 = 18.50$~mag would derive instead. 

As in \citet{mcea04}, we employed four sets of evolutionary tracks to 
compute our HB simulations. These tracks were  
computed by \citet{mcea98} for $Z=0.0005$ and $Z=0.0010$, and by 
\citet{sc98} for $Z=0.0020$ and $Z=0.0060$. The evolutionary 
tracks assume a main-sequence helium abundance of $Y_{\rm MS} = 23\%$ by 
mass and scaled-solar compositions. Helium-enhanced tracks were also computed. 
The mass distribution along the HB in our simulations is represented 
by a normal deviate with a mass dispersion $\sigma_{M}=0.02\,M_{\odot}$. 
In the present paper, we have incorporated color transformation and bolometric 
correction tables for the \emph{uvby} system, as provided by \citet{jcea04}, to 
carry out the transformations from the theoretical $(\log L,\log T_{\rm eff})$ plane 
to the empirical ones. Those tables include the relevant ranges of effective temperature  
and surface gravities for HB stars, thus being clearly adequate for RRL work. We warn 
the reader that the quoted bolometric corrections, over the full range of interest for 
RRL work, differ systematically (${\rm BC}_y = {\rm BC}_V + 0.03$) from those 
adopted in \citet{mcea04}, which in turn came from VandenBerg (1999, priv. comm.). 
This is likely due to a different bolometric correction for the Sun, as adopted in 
the VandenBerg and in the \citeauthor{jcea04} studies. 
Therefore, when compared with the \citet{mcea04} $V$-band magnitudes, the present 
$y$-band magnitudes are, for the same gravity and temperature combination, 
0.03~mag brighter. This also implies distances to the LMC that are correspondingly 
longer, compared to \citet{mcea04} (see \S\ref{sec:STAND} below).   


To define the blue edge of the instability strip, we use equation~(1) in 
\citet{fcea87}, but applying a shift by $-200$~K to the temperature values thus 
derived. The width of the instability strip has been taken as 
$\Delta\log T_{\rm eff}=0.075$; this provides the temperature of the 
red edge of the instability strip for each star once its blue edge has been 
determined. These choices provide a good agreement with more recent theoretical 
prescriptions and the observations \citep[see \S6 in][ for a detailed discussion]{mc04}. 

We include both RRab (also classified as RR0) and RRc (RR1) 
variables in our synthetic PLpsC and PCpsC 
relations. The computed periods for those stars are based on equation~(4) in 
\citet{fcea98}. Therefore, to compare our model prescriptions with 
the observations, the observed RRc periods must be ``fundamentalized'' by 
adding 0.128 to the logarithm of the period.

\section{Genesis of the RRL PLpsC and PCpsC Relations}\label{sec:GENESIS}
In Figures~\ref{fig:GENMAG} and \ref{fig:GENCOL}, we show results for an HB 
simulation computed for $Y_{\rm MS} = 23\%$ and a metallicity $Z = 0.002$, 
and an intermediate HB morphology, as indicated by a value of the Lee-Zinn parameter  
$\mathcal{L} \equiv \mathcal{(B-R)/(B+V+R)} = 0.0$ (where 
$\mathcal{B,R,V}$ represent the numbers of blue, red, and variable~-- RRL-type~-- 
HB stars, respectively). In the {\em upper row} of Figure~\ref{fig:GENMAG}, 
from left to right, one finds the resulting 
color-magnitude diagrams (CMDs) in the $[M_u,\,(b\!-\!y)_0$], $[M_v,\,(b\!-\!y)_0$], 
$[M_b,\,(b\!-\!y)_0$], and $[M_y,\,(b\!-\!y)_0]$ planes, respectively, whereas 
the middle and bottom rows show the resulting PL and (simple, log-linear) PLpsC 
relations. In the upper row of Figure~\ref{fig:GENCOL}, on the other hand, one finds the resulting 
$[(u\!-\!v)_0, C_0]$, $[(v\!-\!b)_0, C_0]$, and $[(b\!-\!y)_0, C_0]$ color-(pseudo-) 
color diagrams, whereas the middle and bottom rows of the same figure show the 
resulting period-color and (simple, log-linear) PCpsC relations. 

In the Str\"omgren system, the magnitudes in the \emph{y} passband are required 
to closely match those in the $V$ passband of the Johnson-Cousins system, which 
is why the CMD in the upper right panel of Figure~\ref{fig:GENMAG} appears so 
similar to those routinely derived in the latter system. Such a similarity also 
allows us to compare our simulations with \citet{mcea04}, since
the latter have shown that the flatness of the HB in their $V$-band simulations 
impacts directly the corresponding PL relation. 
Indeed, as was the case in $V$, the horizontal nature of the 
$M_y,\,(b\!-\!y)_0$ CMD also leads to an essentially flat PL relation in $M_y$, 
with much scatter as a consequence of evolutionary effects.  
In like vein, 
the bluer Str\"omgren passbands show a behavior totally analogous to that 
described in \citet{mcea04} for the bluer $UB$ passbands of the 
Johnson-Cousins system: as can be seen from Figure~\ref{fig:GENMAG}, the effects of   
luminosity and temperature variations are nearly {\em orthogonal} in the 
period-absolute magnitude plane, which leads to large amounts of 
scatter in the corresponding PL relations.

In this sense, \citet{mcea04} have pointed out that such scatter 
is so large in the $UBV$ passbands as to virtually render the corresponding 
PL relations useless, as opposed to what 
happens for redder passbands where the effects of temperature and luminosity 
variations become increasingly parallel in the period-absolute magnitude plane. 
As a consequence, increasingly tight, and therefore useful, PL relations 
obtain towards the near-infrared. In the Str\"omgren system, on the other 
hand, near-infrared bandpasses are lacking, so that one must resort to 
a different strategy in order to obtain useful relationships involving
the pulsation period and the absolute magnitude in this filter system.  

In this sense, our simulations reveal that, if a pseudo-color $C_{0}$ term 
is added to the PL relations, the scatter
essentially disappears (see the bottom row in Fig.~\ref{fig:GENMAG}). This is especially 
encouraging in view of the fact that the  pseudo-color $C_{0}$ of the 
Str\"omgren system is barely affected by reddening 
\citep[e.g.,][]{cm76}, thus being very close to its uncorrected value, 
$C_{1} \equiv (u\!-\!v)-(v\!-\!b)$. Quantitatively, one finds 
$C_{0} = C_{1}-0.15\,E(\bv)$. As a 
consequence, calibrated PLpsC relationships in which this 
pseudo-color index is used are expected to provide us with very 
useful tools to derive the distances to RRL-rich stellar systems, and 
even to individual (e.g., field) RRL stars. 

In like vein, the {\em bottom row} of Figure~\ref{fig:GENCOL} reveals that, by 
adding in a pseudo-color term, one is also able to reduce dramatically the 
scatter that was present in the original period-color relations
({\em middle row} in Fig.~\ref{fig:GENCOL}). Accordingly, our 
resulting PCpsC relationships are also expected to allow one to determine the 
reddenings of the systems to which the RRL belong, based only on the observed 
periods, (reddening-insensitive) pseudo-colors, and the {\em reddened} colors 
of individual RRL stars.

\section{The RRL PLpsC and PCpsC Relations Calibrated}\label{sec:CALIBRE}
We have computed, for each of the four metallicity values indicated in 
\S\ref{sec:MODELS} and a $Y_{\rm MS} = 23\%$, 
extensive sequences of HB simulations that produce 
from very blue to very red HB types. These simulations do not include such 
effects as HB bimodality or the impact of second parameters other than 
age or mass loss on the red giant branch (but see \S\ref{sec:HELIUM} 
below for the effects of an increase in the helium abundance). 

The whole set of simulations contains a total of 336,576 synthetic RRL stars 
with metallicities ranging from $Z = 0.0005$ to $0.006$ and HB morphologies 
from $\mathcal{L} = -0.95$ to +0.95. As 
in \citet{cc08}, we find that, for each individual metallicity, the 
derived Str\"omgren magnitudes are well described by analytical fits 
that involve up to cubic terms in (the log of) $C_0$, whereas a linear 
term in $\log P$ suffices.\footnote{We have experimented with both $C_0$
and $\ln C_0$, and obtained, as a rule, tighter relations when using the 
latter.} 
Rather than providing the forms of these relations for each metallicity 
separately \citep[see][ for the $Z=0.002$ case]{cc08}, 
we have here attempted to obtain fits in which 
all metallicities are simultaneously taken into account. 
This was accomplished by adopting a simple quadratic dependence 
on metallicity for each of the coefficients in the original fit. 
Here we provide fits 
for Str\"omgren $y$ and for the colors $b-y$, $v-b$, and $u-v$, from 
which the individual magnitudes in $u$, $v$, $b$ can be straightforwardly 
computed. 

The final relations that we obtained are thus of the form: 

\begin{eqnarray}
{\rm mag \,\, or \,\,color} = \sum_{i=0}^{2} a_i (\log Z)^i & + & \sum_{i=0}^{2} b_i (\log Z)^i (\ln C_0)   \nonumber \\
															& + & \sum_{i=0}^{2} c_i (\log Z)^i (\ln C_0)^2 \nonumber \\
															& + & \sum_{i=0}^{2} d_i (\log Z)^i (\ln C_0)^3 \nonumber \\
															& + & e_0 (\log P), 
\label{eq:FITS}
\end{eqnarray}

\noindent where ``mag'' stands for the absolute magnitude in $y$, whereas
``color'' represents  
any of the colors $b-y$, $v-b$, and $u-v$. In this expression, $C_{0}$ 
is the Str\"omgren system's pseudo-color, and $P$ is the 
fundamental RRL period (in days). The corresponding coefficients,  
along with their errors, are given in Table~\ref{tab:COEF}. (Naturally, 
the $c_0$ coefficient that appears in this table is {\em not} the same 
as the Str\"omgren pseudo-color $C_0$, which is given in capital letters 
throughout this paper to avoid confusion.)

\begin{deluxetable}{lccclcc}
\tabletypesize{\footnotesize}
\tablecaption{Coefficients of the Fits}
\tablewidth{0pt}
\tablehead{
\colhead{Coefficient} & \colhead{Value} &  \colhead{Error} &&
\colhead{Coefficient} & \colhead{Value} &  \colhead{Error}}
\startdata
\multicolumn{3}{c}{$y$} && \multicolumn{3}{c}{$b-y$} \\ 
\cline{1-3}
\cline{5-7}
$a_0$  & $+0.0895$ & $0.0052$  && $a_0$ & $+0.2440$ & $0.0005$  \\
$a_1$  & $+0.0037$ & $0.0037$  && $a_1$ & $+0.0044$ & $0.0004$  \\
$a_2$  & $-0.0252$ & $0.0007$  && $a_2$ & $-0.0002$ & $0.0001$  \\
$b_0$  & $-0.5242$ & $0.0766$  && $b_0$ & $-0.5472$ & $0.0076$  \\
$b_1$  & $+0.6885$ & $0.0564$  && $b_1$ & $-0.1280$ & $0.0056$  \\
$b_2$  & $+0.1458$ & $0.0102$  && $b_2$ & $-0.0164$ & $0.0010$  \\  
$c_0$  & $+3.2354$ & $0.6837$  && $c_0$ & $-2.3427$ & $0.0682$  \\
$c_1$  & $+3.0981$ & $0.4920$  && $c_1$ & $-1.6363$ & $0.0491$  \\
$c_2$  & $+0.6059$ & $0.0874$  && $c_2$ & $-0.3065$ & $0.0087$  \\  
$d_0$  &$-11.7261$ & $1.5383$  && $d_0$ & $-8.4066$ & $0.1534$  \\
$d_1$  & $-5.5043$ & $1.0822$  && $d_1$ & $-5.7964$ & $0.1079$  \\
$d_2$  & $-0.6123$ & $0.1885$  && $d_2$ & $-1.0051$ & $0.0188$  \\  
$e_0$  & $-2.0066$ & $0.0015$  && $e_0$ & $+0.1044$ & $0.0001$  \\
\cline{1-3}
\cline{5-7}
\multicolumn{3}{c}{$v-b$} && \multicolumn{3}{c}{$u-v$} \\ 
\cline{1-3}
\cline{5-7}
$a_0$  & $+0.5569$ & $0.0007$  && $a_0$ & $+1.5574$ & $0.0007$  \\
$a_1$  & $+0.1121$ & $0.0005$  && $a_1$ & $+0.1124$ & $0.0005$  \\
$a_2$  & $+0.0124$ & $0.0001$  && $a_2$ & $+0.0125$ & $0.0001$  \\
$b_0$  & $-0.5220$ & $0.0103$  && $b_0$ & $+0.4855$ & $0.0104$  \\
$b_1$  & $-0.1267$ & $0.0076$  && $b_1$ & $-0.1215$ & $0.0077$  \\
$b_2$  & $-0.0087$ & $0.0014$  && $b_2$ & $-0.0078$ & $0.0014$  \\  
$c_0$  & $-1.5715$ & $0.0918$  && $c_0$ & $-1.0571$ & $0.0930$  \\
$c_1$  & $-1.0406$ & $0.0661$  && $c_1$ & $-1.0305$ & $0.0669$  \\
$c_2$  & $-0.1842$ & $0.0117$  && $c_2$ & $-0.1830$ & $0.0119$  \\  
$d_0$  & $-7.6106$ & $0.2067$  && $d_0$ & $-7.4495$ & $0.2093$  \\
$d_1$  & $-4.8869$ & $0.1454$  && $d_1$ & $-4.8832$ & $0.1472$  \\
$d_2$  & $-0.7979$ & $0.0253$  && $d_2$ & $-0.7988$ & $0.0256$  \\  
$e_0$  & $+0.0882$ & $0.0002$  && $e_0$ & $+0.0884$ & $0.0002$  \\
\enddata
\label{tab:COEF}
\end{deluxetable}

\begin{deluxetable}{lccclcc}
\tabletypesize{\footnotesize}
\tablecaption{Analytical Fits: Quality Diagnostics}
\tablewidth{0pt}
\tablehead{
\colhead{Fit} & \colhead{$r$} &  \colhead{std. error}}
\startdata
$y$    & $+0.9978$ & $0.0084$  \\
$b-y$  & $+0.9998$ & $0.0008$  \\
$v-b$  & $+0.9996$ & $0.0011$  \\
$u-v$  & $+0.9999$ & $0.0011$  \\
\enddata
\label{tab:QUAL}
\end{deluxetable}

\begin{figure*}
 \centering
  \includegraphics*[width=13.5cm]{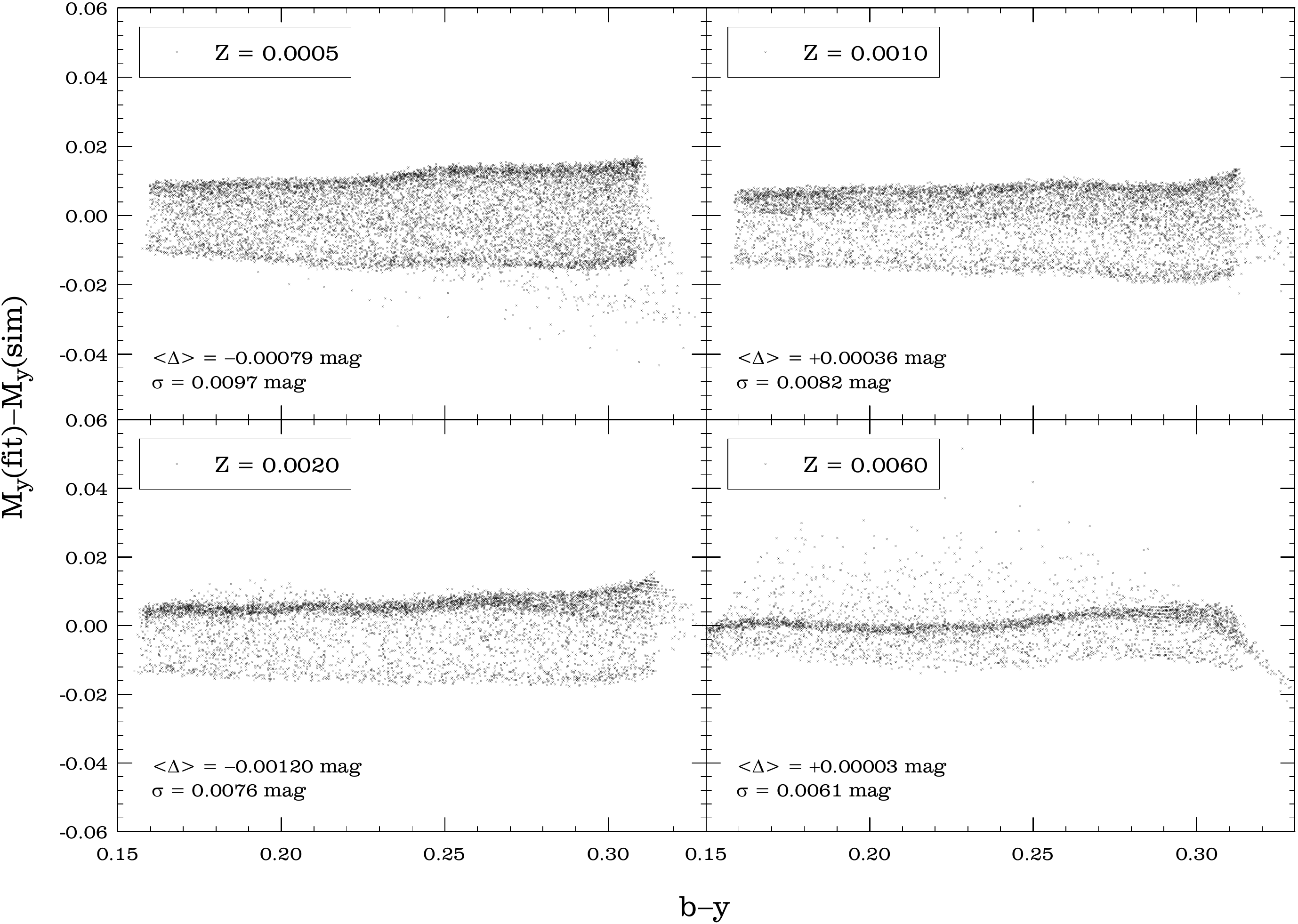}
  \caption{Difference between the absolute magnitude in $y$, as predicted by 
    equation~(\ref{eq:FITS}), and its input value (from the HB simulations), 
    plotted as a function of the $b-y$ color (from the same simulations)  
    for four different metallicities: $Z = 0.0005$ ({\em upper left}), 0.001
    ({\em upper right}), 0.002 ({\em lower left}), 0.006 ({\em lower right}). 
	A total of 28,992 randomly selected synthetic stars is shown for all metallicities. 
    The average magnitude difference is indicated in the lower left corner of
	each panel, along with the corresponding standard deviation. The
    latter is a direct indicator of the precision with which equation~(\ref{eq:FITS})
    is able to provide the $M_y$ values. Note, accordingly, that for all 
    metallicities equation~(\ref{eq:FITS}) yields $M_y$ values with a precision
    better than 0.01~mag. 
    }
      \label{fig:RESy}
\end{figure*}

\begin{figure*}
 \centering
  \includegraphics*[width=13.5cm]{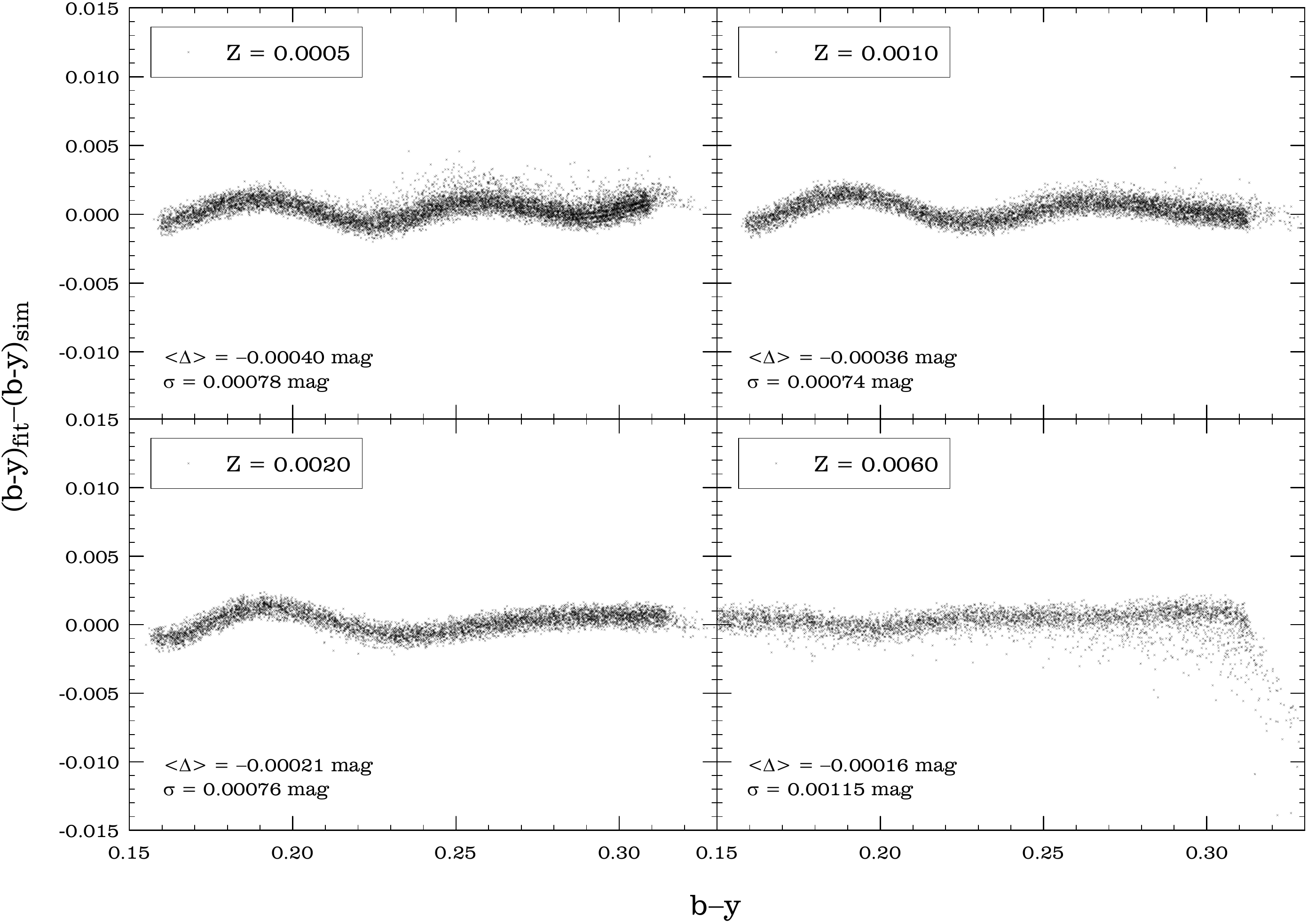}
  \caption{As in Figure~\ref{fig:RESy}, but for $b-y$. Note that the 
    precision in the $b-y$ colors provided by equation~(\ref{eq:FITS}) is at 
    the mmag level. 
    }
      \label{fig:RESbmy}
\end{figure*}

\begin{figure*}
\centering
  \includegraphics*[width=13.5cm]{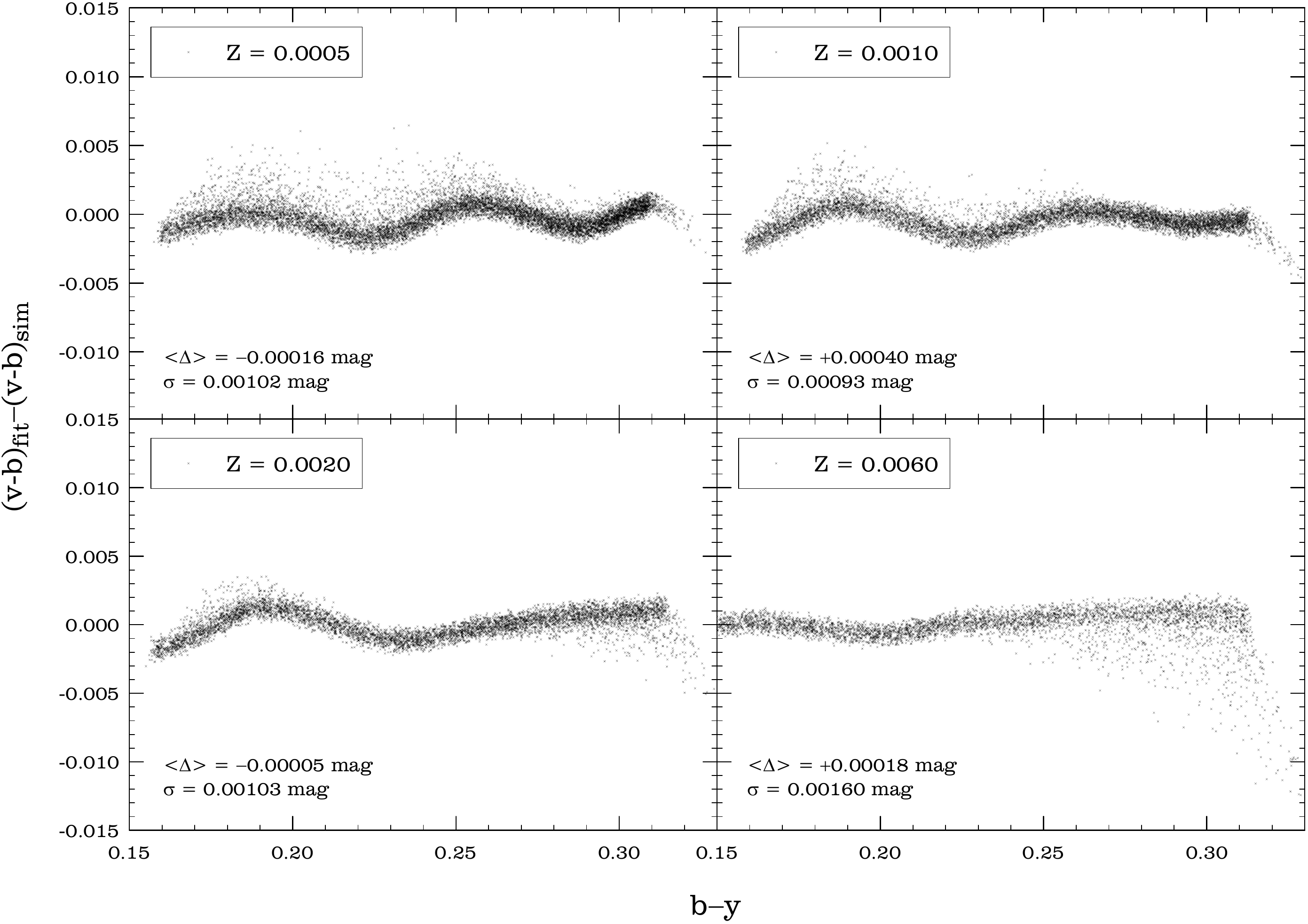}
  \caption{As in Figure~\ref{fig:RESy}, but for $v-b$. Note that the 
    precision in the $v-b$ colors provided by equation~(\ref{eq:FITS}) is at 
    the mmag level.
    }
      \label{fig:RESvmb}
\end{figure*}

\begin{figure*}
\centering
  \includegraphics*[width=13.5cm]{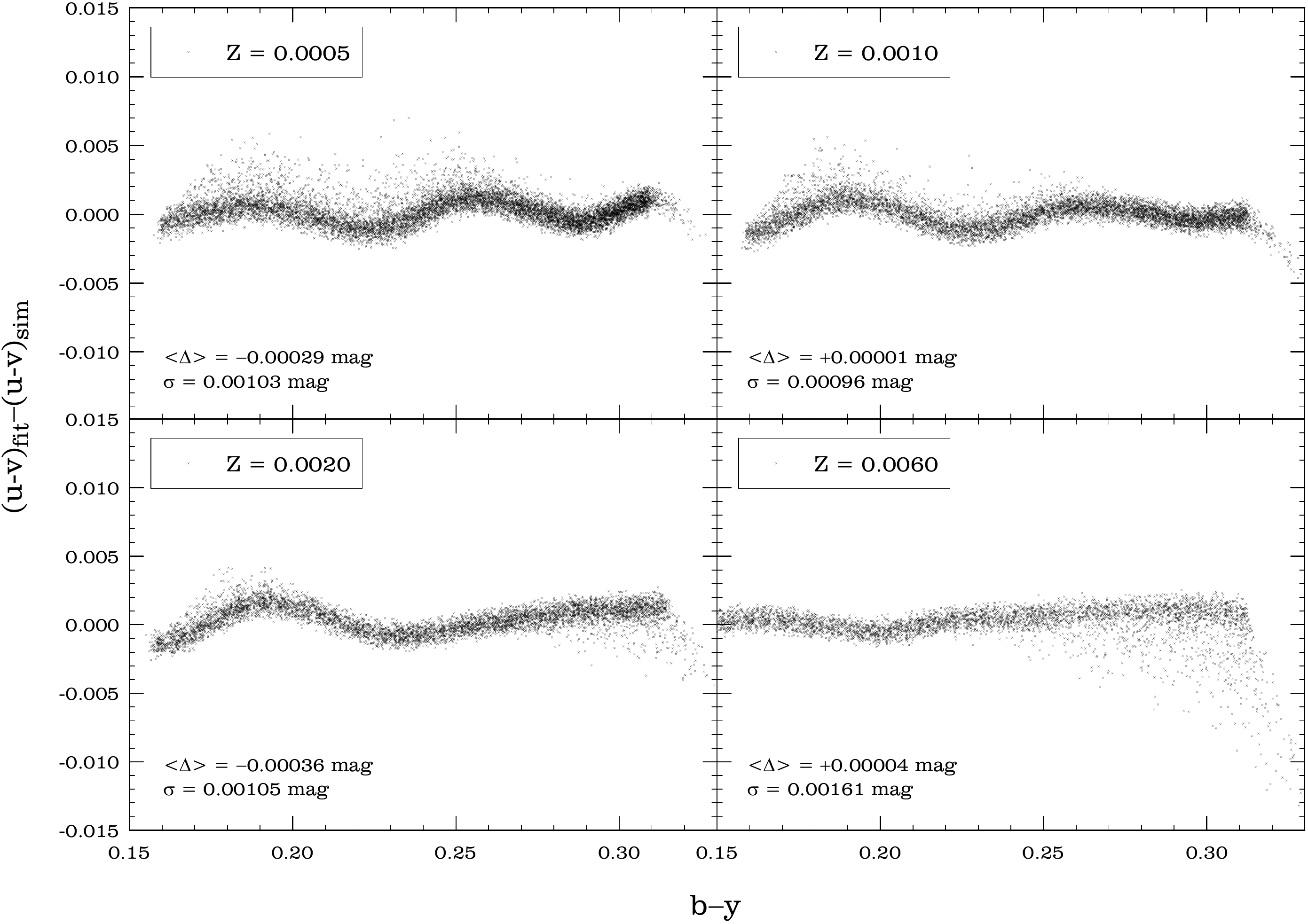}
  \caption{As in Figure~\ref{fig:RESy}, but for $u-v$. Note that the 
    precision in the $u-v$ colors provided by equation~(\ref{eq:FITS}) is at 
    the mmag level.
  }
      \label{fig:RESumv}
\end{figure*}

\begin{figure*}
 \centering
  \includegraphics*[width=13.5cm]{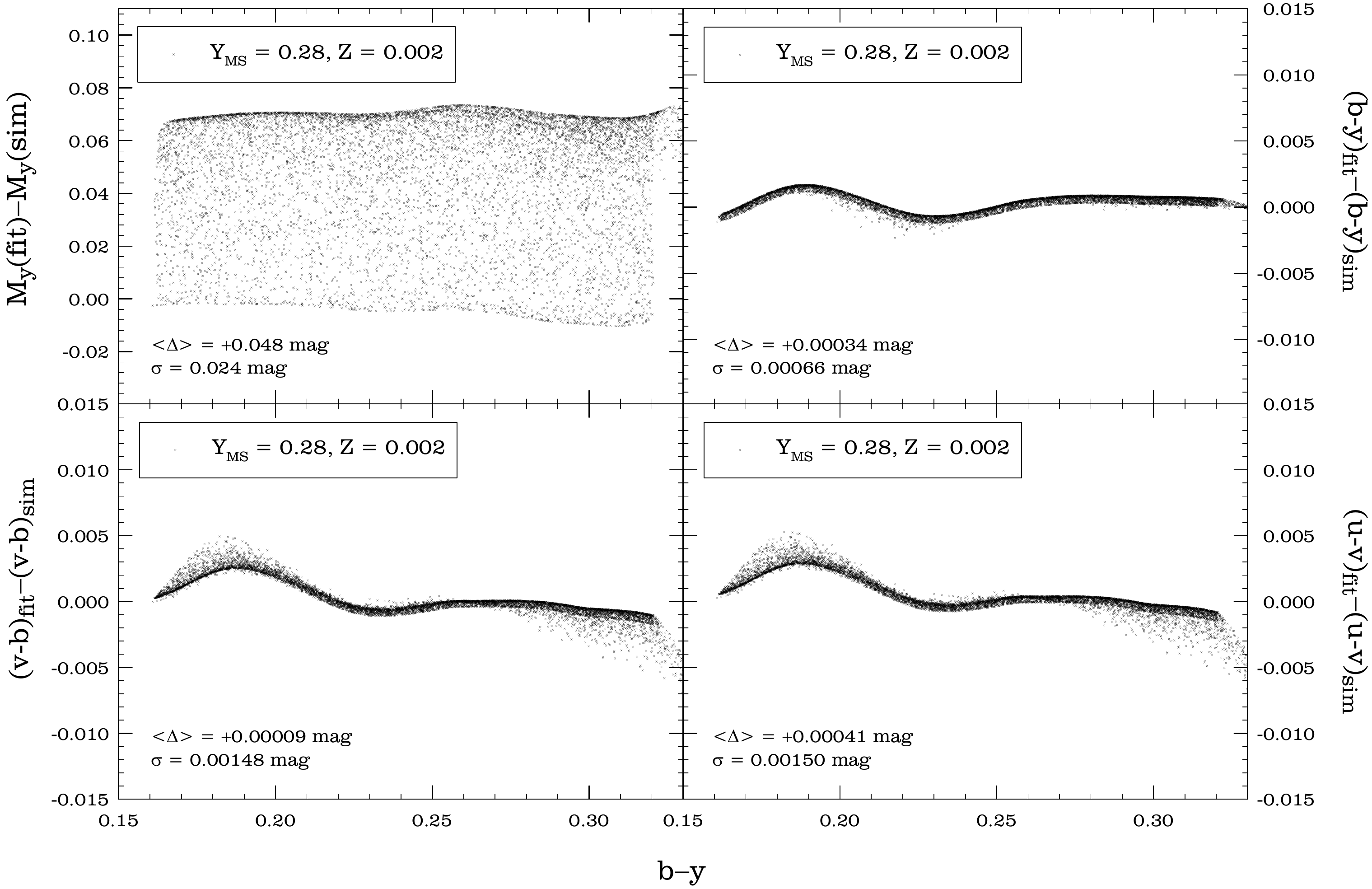}
  \caption{Effect of an increase in the helium abundance upon the derived 
    relations. Here we give the difference between the quantities predicted 
    by equation~(\ref{eq:FITS}) and its input value from HB simulations computed 
    for a $Y_{\rm MS} = 0.28$ and $Z = 0.002$. {\em Upper left:} $M_y$; 
    {\em upper right:} $b-y$; {\em lower left:} $v-b$; {\em lower right:} $u-v$. 
    All quantities are plotted as a function of the $b-y$ value from the 
	simulations. A total of 8338 synthetic stars is shown in each panel. 
	As can be seen, an increased helium has a clear influence upon the absolute 
    magnitudes, but not so upon the derived colors.     
    }
      \label{fig:HIY}
\end{figure*}

We stress that these equations are able to reproduce the input values 
(from the HB simulations) with remarkable precision. This is shown in 
Table~\ref{tab:QUAL}, where the correlation coefficient $r$ and the 
standard error of the estimate are given for each of the four 
equations. We also show, in Figures~\ref{fig:RESy}, \ref{fig:RESbmy}, 
\ref{fig:RESvmb}, and \ref{fig:RESumv}, the residuals [in the sense 
eq.~(\ref{eq:FITS}) minus input values (from the simulations)] 
for a random subset of 28,992 synthetic stars drawn from the original pool of 
336,576 synthetic RRL stars in the HB simulations, for the fits computed for 
$y$, $b-y$, $v-b$, and $u-v$, respectively. These plots further illustrate  
that the Str\"omgren magnitudes and colors can be predicted from the data 
provided in equation~(\ref{eq:FITS}) and Table~\ref{tab:COEF} with great 
precision. 

Finally, we note that equation~(\ref{eq:FITS}) can be trivially expressed
in terms of [Fe/H]; this can be accomplished using the relation 

\begin{equation}
  \log Z = {\rm [M/H]} - 1.765, 
\label{eq:LOGZ}
\end{equation} 

\noindent which is the same as equation~(9) in \citet{mcea04}. In this 
sense, the effects of an enhancement in $\alpha$-capture elements 
with respect to a solar-scaled mixture, such as observed amongst 
Galactic halo stars \citep*[e.g.,][ and references therein]{bpea05b}, 
can be taken into account by using the following scaling relation 
\citep*{msea93}: 

\begin{equation}
{\rm [M/H]} = {\rm [Fe/H]} + \log (0.638\,f + 0.362), 
\label{eq:MAUR}
\end{equation} 

\noindent where $f = 10^{\rm [\alpha/Fe]}$. However, such a relation 
should be used with due care for metallicities $Z > 0.003$ \citep{dvea00}.

\subsection{The Effect of Helium Abundance}\label{sec:HELIUM} 
As already stated (\S\ref{sec:MODELS}), the aforementioned simulations are 
based on a fixed mass dispersion ($\sigma_M = 0.020\,M_{\odot}$) and helium 
abundance (23\%). However, several authors have pointed out that old, 
helium-enhanced populations may exist, even at low metallicities, thus 
opening the possibility that helium-enhanced RRL stars may also exist  
\citep[see][ for a review]{mc05}. How would a helium enhancement affect 
the derived relations for RRL stars? 

In order to answer this question, 
we have computed additional sets of synthetic HBs for a 
$Y_{\rm MS} = 28\%$ ($Z=0.0020$), 
and applied the same relations as in equation~(\ref{eq:FITS}) to check how 
much we would err by assuming that the relations derived for a 
$Y_{\rm MS} = 23\%$ are 
valid also for a higher $Y$. The enhanced-$Y$ simulations cover the full 
range of HB types, from very blue to very red, and include a total of 8338 
synthetic RRL stars. 

The residuals, in the sense equation~(\ref{eq:FITS}) minus simulations, are 
shown in Figure~\ref{fig:HIY}. In this figure, the upper left panel shows the 
residuals for $M_y$, the upper right one for $b-y$, the lower left one for 
$v-b$, and the lower right one for $u-v$. As in the previous figures, the 
average residual and standard deviation are also provided in the insets.  

It is immediately apparent that our derived relations for $b-y$, $v-b$, and 
$u-v$ can be safely applied to RRL stars with a significantly enhanced helium 
content, the implied errors being significantly smaller than 0.01~mag. 
In the case of $M_y$, there is a clear offset in the zero point, as 
well as an increased dispersion in comparison with Figure~\ref{fig:RESy}. 
Still, the standard deviation remains a modest $0.024$~mag, and the full 
dispersion range is of order 0.08~mag only~-- to be compared with the actual 
dispersion in $M_y$ magnitudes from the HB simulations in the enhanced-$Y$
case, which amounts to a full 0.59~mag (i.e., $M_y$ values for individual 
RRL stars range from 0.58~mag at their faintest to $-0.01$~mag at their brightest). 
Therefore, if a correction to the zero point $a_0$ for $M_y$ in Table~\ref{tab:COEF}
by $dM_y/dY = -0.048/(0.28-0.23) = -0.96$ 
(i.e., in the sense that eq.~\ref{eq:FITS} predicts too faint magnitudes) is 
duly taken into account, equation~(\ref{eq:FITS}) can also be used to provide 
useful information on the absolute magnitudes of RRL stars with enhanced  
helium abundances.  

To close, we note that, as in \citet{mcea04}, we have also checked the effect 
of changing the mass dispersion $\sigma_M$ upon our results, and found that 
equation~(\ref{eq:FITS}) adequately covers the possibility of different
$\sigma_M$ values.

\subsection{Average Relations for $M_y$}\label{sec:STAND}
It is common practice, in RRL work, to employ a simple, linear 
relation between {\em average} absolute magnitude in the visual and 
[Fe/H] \citep[e.g.,][ and references therein]{mc05}. 
On the basis of our models, we find the following relation: 

\begin{equation}
M_y = (1.4350\pm 0.0009) + (0.2800\pm 0.0003)\, \log Z, 
\label{eq:LIN} 			  
\end{equation}

\noindent with a correlation coefficient $r = 0.835$. 
As shown by several authors, and as also discussed by Catelan et al. 
(2004), a quadratic relation may be superior to such a simple linear relation; 
we accordingly obtain: 

\begin{eqnarray}
M_y  =  (2.259\pm 0.007)  & + & (0.883\pm 0.005) \, \log Z \nonumber \\
			  & + & (0.108\pm 0.001)\, (\log Z)^{2}, 
\label{eq:QUAD} 			  
\end{eqnarray}

\noindent with a correlation coefficient $r = 0.843$. (The small errors in 
the coefficients are a consequence of the large number of synthetic stars used in 
deriving these relations.) Note that this expression is basically identical to 
equation~(8) in \citet{mcea04}, which was obtained for the $V$ band following
a similar procedure as in the present paper, except that their zero points 
differ by 0.03~mag (in the sense that eq.~\ref{eq:QUAD} provides slightly 
brighter magnitudes, as expected; see \S\ref{sec:MODELS}). 
The true distance modulus to the LMC 
implied by equation~(\ref{eq:QUAD}) is 18.49~mag 
\citep[using the data for LMC RRL from][]{rgea04} or 18.53~mag 
\citep[using instead data from][]{jbea04}. 
Such a value is in agreement, within the  
errors, with the distance modulus recently determined by \citet{cc08}, who 
pegged the zero point of their distance scale to RR Lyr's trigonometric 
parallax, and found $(m-M)_0^{\rm LMC} = 18.44 \pm 0.11$. 

\begin{figure}
\centering
  \includegraphics*[width=8cm]{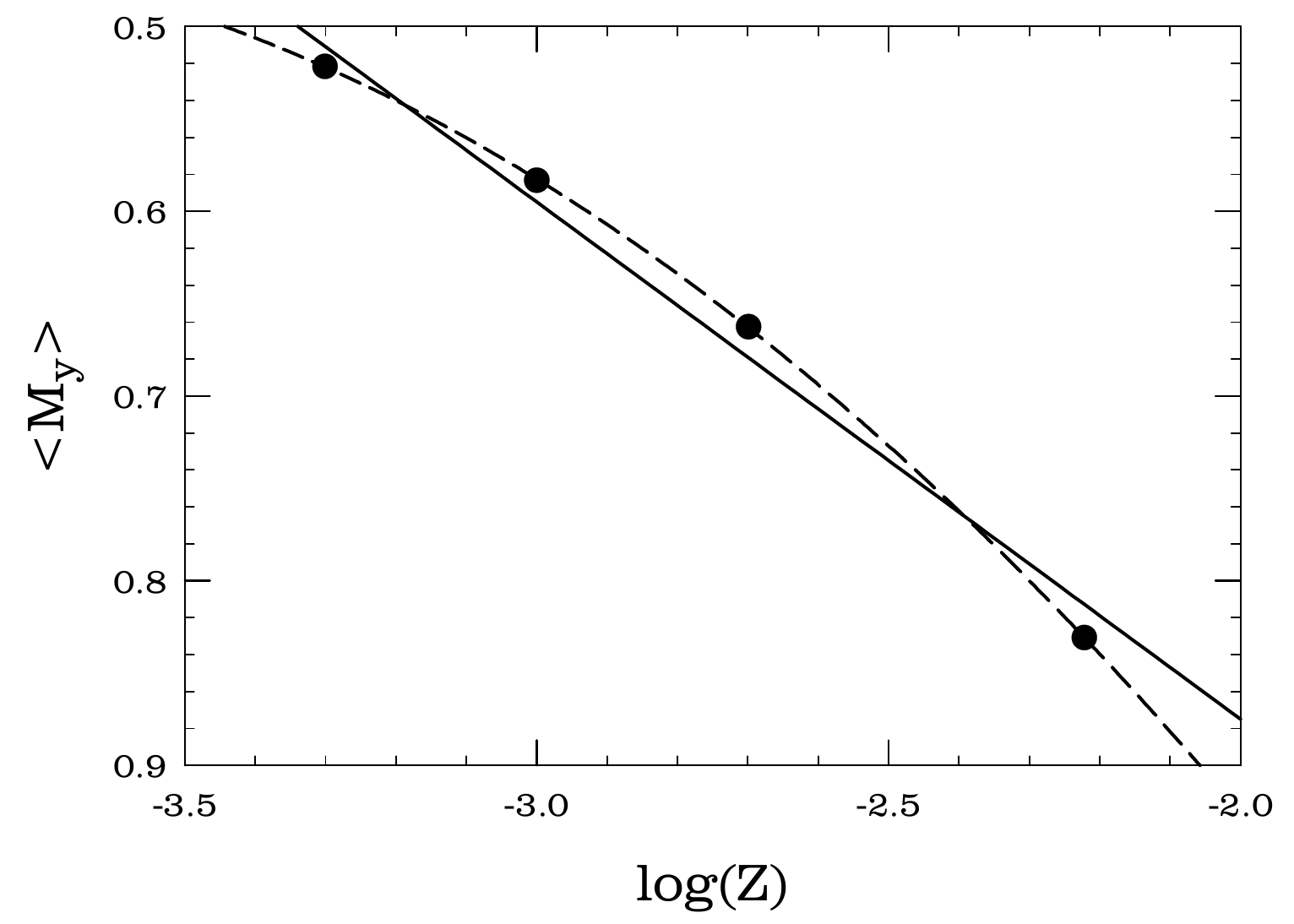}
  \caption{Correlation between average $M_y$ magnitude and metallicity, 
    with equations~(\ref{eq:LIN}) and (\ref{eq:QUAD}) ({\em continuous} and 
	{\em dashed} lines, respectively) overplotted.
  }
      \label{fig:avMy}
\end{figure}

Equations~(\ref{eq:LIN}) and (\ref{eq:QUAD}) are 
shown in Figure~\ref{fig:avMy}. Naturally, absolute magnitudes for individual 
RRL stars, as derived on the basis of either of these equations, will be 
much less reliable than those based on equation~(\ref{eq:FITS}), since the 
latter is the only one that is able to take evolutionary effects into 
account {\em on a star-by-star basis}. 
We will come back to this point momentarily \citep[see also][]{cc08}.

\section{On Applying Our Relations to RRL Stars}\label{sec:CAVEATS} 
When applying our equation~(\ref{eq:FITS}) in globular cluster work, 
the metallicity of the cluster will often be known a priori. However, 
metallicity estimates may also be unavailable, especially when dealing 
with field RRL stars. Yet, for a reliable application of our relations 
to field stars, an estimate of their metallicities must be provided. 

The Str\"omgren system itself may again come to our rescue in such a case. 
In a forthcoming paper, we shall provide a technique to derive metallicities 
for RRL stars, based on the Str\"omgren parameters $m_0 = (v-b)_0 - (b-y)_0$ 
(also called the Str\"omgren ``metal-line index''; \citeauthor{bs63} 
\citeyear{bs63}) and $C_0$. In the meantime, we recall that RRL 
metallicities can also be obtained on the basis of their $V$-band light curves 
using Fourier decomposition \citep*[e.g.,][]{jk96,jj98,kk07,smea07}; 
since the $y$-band magnitudes are very 
similar to those in the Johnson-Cousins $V$-band, this means that available 
calibrations of Fourier decomposition parameters as a function of metallicity 
can be employed in the Str\"omgren system as well.\footnote{We have checked 
that this statement is indeed correct in the case of the RRL star X~Ari, 
which was studied in both photometric systems by 
\citet{rjea87}: not 
only do X~Ari's light curves in $V$ and $y$ look very similar, but also the 
low- (i.e., up to 5th) order Fourier parameters derived therefrom are 
in very good agreement. As a result, the star's corresponding Fourier-based 
metallicities, as derived from the $V$- or $y$-band light curves, are within 
0.1~dex of one another.}  
Naturally, the reader should keep in mind that such calibrations can 
only be applied to variable stars that do not show the \citet{sb07} effect. In 
addition, in \citeauthor{jk96} a deviation parameter $D_{m}$ was defined to 
measure the degree of reliability of an ab-type RRL light curve for 
application of the Fourier decomposition method. In particular, when 
$D_{m} < 3$, the physical, chemical, and photometric parameters, as derived 
from Fourier fits, should be considered more reliable. Unfortunately, 
high-$S/N$ light curves in the Str\"omgren system are more difficult 
to obtain than in the Johnson-Cousins system, due to the fact that in the 
former case we are 
dealing with an intermediate-band system. Longer exposure times may 
accordingly be needed, but care should be taken to keep these 
exposures sufficiently short that an insignificant fraction of the star's 
pulsation period is encompassed by them.

The reader should be warned that our relations should be compared 
against empirical quantities obtained for the so-called {\em equivalent static 
star}. Several procedures have been advanced in the literature for the 
determination of the latter on the basis of empirically derived magnitudes 
and colors 
\citep*[e.g.,][ and references therein]{gbea95}. In particular, one should note 
that, according to the hydrodynamical models provided by \citeauthor{gbea95}, 
one should expect differences between temperatures derived from  
intensity- or magnitude-averaged colors, on the one hand, and those based 
on the actual color of the equivalent static star, on the other. As a workaround, 
these authors set forth very useful amplitude-dependent corrections, which become 
more important the bluer the bandpass. Unfortunately, to the best of our 
knowledge a similar study has never been extended to bandpasses bluer than 
$B$ \citep*[but see][]{gbea94}, as would be required to 
quantitatively evaluate the impact of nonlinear phenomena upon the derived 
average quantities using $u$ and $v$ in particular. In this sense, 
an extension to the Str\"omgren system of the work 
that was carried out by Bono et al. in the Johnson-Cousins system would 
be of great interest. On the other hand, and as 
noted by the referee, $C_0$, being a difference between two colors, 
is presumably affected to a lesser degree than are the colors themselves
\citep[see also][]{cc08}.

\begin{deluxetable*}{lccccccccc}
\tabletypesize{\footnotesize}
\tablecaption{Applications to Field RRL Stars}
\tablewidth{0pt}
\tablehead{
\colhead{Star} &\colhead{$P$ (day)} &  \colhead{[Fe/H]} &  \colhead{$C_{0}$} &\colhead{$(b-y)_{0}$\tablenotemark{(a)}}  
& \colhead{$(b-y)_{0}$\tablenotemark{(b)}} & \colhead{$\Delta(b-y)_{0}$}
& \colhead{$M_y$\tablenotemark{(c)}}& \colhead{$M_y$\tablenotemark{(d)}} & \colhead{$\Delta M_{y}$} }
\startdata
SV Hya	&	0.479	&	$-1.10$	&	0.903	&	0.212	& 0.229 & 0.017 &	0.680	&	0.674	& $-0.006$	\\
RR Lyr\tablenotemark{(e)}	
	&	0.567	&	$-1.39$\tablenotemark{(f)}
					&	0.853	& 	0.241	& 0.249 & 0.008 &	0.599	&	0.549	& $-0.051$	\\
RR Lyr\tablenotemark{(e)}	
	&	0.567	&	$-1.16$\tablenotemark{(g)}	
					&	0.853	& 	0.241	& 0.252 & 0.011 &	0.662	&	0.587	& $-0.075$	\\
SU Dra	&	0.661	&	$-1.38$	&	0.864	&	0.252	& 0.256 & 0.008 &	0.602	&	0.402	& $-0.200$	\\
SS Leo	&	0.627	&	$-1.50$	&	0.837	&	0.246	& 0.257 & 0.011 &	0.573	&	0.463	& $-0.111$	
\enddata
\tablenotetext{a}{From \citet{dm97}.}
\tablenotetext{b}{Using  equation~(\ref{eq:FITS}).}
\tablenotetext{c}{Using  equation~(\ref{eq:QUAD}).}
\tablenotetext{d}{Using  equation~(\ref{eq:FITS}).}
\tablenotetext{e}{Parameters from \citet{cc08}.}
\tablenotetext{f}{From \citet{gcea95}, in the \citet{zw84} scale.}
\tablenotetext{g}{From \citet{gcea95}, translated to the \citet{cg97} scale.}
\label{tab:OBS}
\end{deluxetable*}

As a first, rough 
approximation to the amplitude-dependent corrections that may be expected 
in the case of the $(b\!-\!y)_0$ color index, Table~4 in \citet{gbea95}, 
which was computed for the broadband $\bv$ color, may be used as a guide.  
This table indicates that, except at the very highest amplitudes 
($A_B > 1.8$~mag), the corrections needed to go from the average $\bv$ color 
(computed directly in intensity or magnitude units) to the color of the 
equivalent static star are always smaller than 0.025~mag for an ab-type 
RRL star (a similar limit obtains for RRc stars with amplitudes that 
are smaller than $A_B = 1$~mag). It is worth noting, in this sense, that 
\citet{cc08}, in their study of the star RR Lyr, have indeed found only 
small differences between the average magnitudes (including the bluer 
filters), colors, and $C_0$ values, 
computed following different recipes for finding the properties of the 
equivalent static star.   
 
Observers are also strongly warned against using single-epoch 
photometry to derive $C_1$ values to be used along with our relations: 
according to the light curves presented by \citet{ms82}, ab-type RRL 
may present amplitudes in $C_1$ that may reach up to 0.6~mag. According 
to our derived relations, changes in $C_0$ by $\pm 0.3$~mag lead to average 
changes 
in $M_y$ by      $^{-0.40\pm 0.03}_{+0.61\pm 0.27}$~mag, 
in $b\!-\!y$ by  $^{-0.10\pm 0.01}_{+0.11\pm 0.05}$~mag, 
in $v\!-\!b$ by  $^{-0.07\pm 0.01}_{+0.12\pm 0.07}$~mag, and 
in $u\!-\!v$ by  $^{+0.23\pm 0.01}_{-0.18\pm 0.07}$~mag. 

To close, we also note that there do exist some calibrations
to derive $T_{\rm eff}$ from $(b\!-\!y)_{0}$, which show that this color  
is indeed a very good indicator of effective temperature 
\citep[e.g.,][]{jcea04}. Unfortunately, 
these calibrations are based on evolutionary states that precede core
helium-burning stars. On the other hand, the plane $(b\!-\!y)_{0}-C_{0}$ has  
been described as a good indicator of effective temperature (and surface 
gravities) for RRL stars \citep[e.g.,][]{ms82,dm97}. Given its tremendous 
potential, it is very unfortunate that there are not more empirical studies 
based on Str\"omgren filters for RRL stars, the papers by \citet{ie69} and 
\citet{ms82} being notable exceptions.

\section{Comparisons with the Observations}\label{sec:DATA}
While a detailed comparison with the empirical data is beyond the scope of the 
present paper, in this section we provide a first application of our relations 
to observations of field RRL stars \citep[see also][]{cc08}. In this sense, 
Table~\ref{tab:OBS} provides a comparison between colors and magnitudes, as 
derived from our relations, and (in the case of the $b-y$ colors) those from 
previous studies \citep[from][ unless otherwise noted]{dm97}, for stars within 
the general range of validity of our relations. 
In column 1, we provide the star's name; in column 2, the period (in days). The 
metallicity [Fe/H] is indicated in column 3, whereas column 4 gives the star's $C_0$ 
value. Column 5 gives the star's $(b\!-\!y)_0$ color from \citet{dm97}, whereas 
column 6 gives the same quantity, as derived from the listed $C_0$ and $P$ values 
on the basis of our equation~(\ref{eq:FITS}). Column 7 gives the difference between 
these two color estimates. 
Column 8 provides $M_y$, based 
on our equation~(\ref{eq:FITS}), whereas column 9 lists the $M_y$ value  
derived on the basis of equation~(\ref{eq:QUAD}). Column 10 lists the difference 
between these two $M_y$ estimates. 

For the star RR Lyr, we provide two different rows, 
corresponding to two different possibilities for the 
star's metallicity value, both based on the \citet{gcea95} measurements  
(which, according to \citeauthor{abea01} \citeyear{abea01}, provide [Fe/H] for 
the star in the \citeauthor{zw84} \citeyear{zw84} scale). The global 
metallicity $Z$ is obtained, on the basis of the indicated [Fe/H] values, 
using equations~(\ref{eq:LOGZ}) and (\ref{eq:MAUR}). An $\alpha$-enhancement 
$[\alpha/{\rm Fe}] = +0.31$ \citep[again from][]{gcea95} is assumed for 
${\rm [Fe/H]} < -1$, and $[\alpha/{\rm Fe}] = 0$ at higher metallicities. 

\subsection{Colors}\label{sec:COLORS}
Comparing the $(b\!-\!y)_0$ colors derived on the basis of our equation~(\ref{eq:FITS}) 
with those from \citet{dm97}, we find an average difference of $0.011$~mag (in the sense 
our fit minus McNamara), with standard deviation $0.005$~mag, and a maximum difference of 
$0.017$~mag (see Table~\ref{tab:OBS}). Clearly, our relations provide a very good match 
to McNamara's intrinsic color determinations for RRL stars, at least in $(b\!-\!y)$, 
though a systematic overestimate by $\approx 0.01$~mag cannot be ruled out. 

\subsection{Absolute Magnitudes}\label{sec:MAGS} 
Note that $M_y$ values, as derived on the basis of equation~(\ref{eq:FITS}), do 
indeed refer to the absolute magnitude {\em of the individual star}. $M_y$ values 
derived on the basis of a simple $M_y-{\rm [Fe/H]}$ relation such as 
equation~(\ref{eq:QUAD}), on the other hand, simply provide the {\em average} 
absolute magnitude of RRL stars of metallicity similar to a given star's. It 
thus follows that a comparison between these two quantities provides us with a 
direct estimate of the {\em degree of overluminosity} (due to evolutionary 
effects) in $y$ (and thus similarly in $V$) of individual RRL stars 
\citep[see also][]{cc08}. This degree of overluminosity is then precisely 
what the last column of Table~\ref{tab:OBS} provides. 

It thus appears that all of the stars that we have studied but one 
(SV Hya) are somewhat overluminous compared to the expected mean $M_y$ for their 
metallicities. In the case of RR Lyr itself, 
we find an overluminosity in the range between $0.051-0.075$~mag, depending 
on the adopted metallicity scale (see Table~\ref{tab:OBS}), 
which is only slightly smaller than the similar 
result (namely, an overluminosity of $0.077\pm 0.010$~mag in $y$) recently 
obtained by \citet{cc08} on the basis of relations that they derived 
for $Z = 0.002$ models. As pointed out by 
\citeauthor{cc08}, due to the fact that empirical versions of our 
equations~(\ref{eq:LIN}) and (\ref{eq:QUAD}) are normally based on the 
{\em assumption} that RR Lyr is representative of the average for its 
metallicity, a correction to the zero points of these empirical calibrations 
by this same amount is required (as indeed performed by \citeauthor{cc08}).
This, and again as shown by \citeauthor{cc08} on the basis of 
the latest trigonometric parallax values for RR Lyr, leads to a revised 
true distance modulus for the Large Magellanic Cloud of 
$(m-M)_0 = 18.44 \pm 0.11$~mag.

\section{Conclusions}\label{sec:CONCL}
We have presented the first calibration of the 
RRL PLpsC and PCpsC relations in the \emph{uvby} bandpasses
of the  Str\"omgren system. Though we have shown that ``pure'' PL and PC 
relations do not exist in this system due to the scatter brought about 
by evolutionary effects, we have also demonstrated that the latter 
can be satisfactorily taken into account by including pseudo-color- (i.e., 
$C_{0}$-) dependent terms in the calibration~-- thus leading to our 
reported period-luminosidy-{\em pseudo}-color (PLpsC)
and period-color-{\em pseudo}-color (PCpsC) relations. The latter 
are provided in the form of analytical fits (eq.~\ref{eq:FITS} and 
Table~\ref{tab:COEF}), which we show to be able to provide $M_y$ 
values that can be trusted at the $\approx \pm 0.01$~mag level, 
and $b-y$, $v-b$, and $u-v$ colors that are good at the mmag level. 
We also show that these relations remain good even in the case of 
helium-enhanced RRL stars, and provide a helium-dependent correction 
to the zero point of the relation for $M_y$. These relations should 
be of great help in deriving reddenings and distances to even 
individual field RRL stars.  

By applying our derived relations to a sample of four field RRL 
stars with Str\"omgren parameters from the literature \citep{dm97}, we 
find evidence that RR Lyr, SU Dra, and SS Leo are overluminous in $y$ (and 
thus $V$) compared to other stars of similar metallicity, by $0.05-0.08$~mag
(depending on the metallicity scale), 0.20~mag, and 0.11~mag, respectively
\citep[see also][]{cc08}. SV Hya, on the other hand, appears more 
representative of the average for its peers, to within 0.01~mag. 
The fact that we can derive the evolutionary status of even individual field 
RRL stars using our relations clearly demonstrates the great potential of 
Str\"omgren photometry in applications of RRL stars to studies of the 
Galactic and extragalactic distance scale.

\acknowledgments We warmly thank an anonymous referee for several 
perceptive comments that led to a significant improvement in the presentation 
of our results. This work is supported by Proyecto FONDECYT Regular 
No. 1071002.

\end{document}